\documentstyle[aps,prl,amsfonts]{revtex}
\begin{document}
\twocolumn
\draft
\title{\bf Topological aspects of geometrical signatures of phase transitions}
\author{Roberto Franzosi$^{1,5,}$\cite{roberto}, Lapo Casetti$^{2,}$\cite{lapo}
, Lionel Spinelli$^{3,5,}$\cite{lionel}, and Marco Pettini$^{4,5,}$\cite{marco}}
\address{$^1$Dipartimento di Fisica, Universit\`a di Firenze, Largo Enrico
 Fermi 2, 50125 Firenze, Italy\\
$^2$ Istituto Nazionale per la Fisica della Materia, Dipartimento di Fisica, 
Politecnico di Torino, \\
Corso Duca degli Abruzzi 24, 10129 Torino, Italy\\
$^3$Centre de Physique Th\'eorique du C.N.R.S., Luminy Case 907,
13288 Marseille Cedex 9, France\\
$^4$Osservatorio Astrofisico di Arcetri, Largo Enrico Fermi 5, 
50125 Firenze, Italy\\
$^5$ Istituto Nazionale per la Fisica della Materia, Unit\`a di Ricerca di 
Firenze, Firenze, Italy} 
\date {\today}
\maketitle
\begin{abstract}
Certain geometric properties of submanifolds of configuration space are 
numerically investigated for classical lattice $\varphi^4$
models in one and two dimensions. Peculiar behaviors of the computed 
geometric quantities are found 
only in the two-dimensional case, when a phase transition is present.
The observed phenomenology strongly supports, though in an indirect way, 
a recently proposed {\it topological conjecture} about a topology change
of the configuration space submanifolds as counterpart of a phase transition.

\end{abstract}
\vskip 1truecm
\pacs{PACS numbers: 05.70.Fh; 64.60.-i; 02.40.-k ; 05.20.-y }
\narrowtext
In Statistical Mechanics phase transitions are associated with 
the appearence of the so-called Yang-Lee (real)zeros \cite{LeeYang} of the 
grand-partition function entailing singular temperature-dependence 
of thermodynamic observables.  
However in the Yang-Lee theory the necessary conditions for the existence  of 
real zeros remain unspecified. 
It has been recently suggested \cite{cccp,cccppg} 
that the thermodynamic singularities might have their
deep origin in some major topological change in configuration space, i.e.
in a non-trivial structure of the {\it support} 
of the equilibrium statistical measure.

This {\it topological conjecture} \cite{nota1} has been put forward 
heuristically 
within the framework of a numerical investigation of the Hamiltonian 
dynamical counterpart of phase transitions. An interesting outcome of
such investigations was the clear evidence of a peculiar temperature-behavior
of the largest Lyapunov exponent at the phase transition point. This was
observed in lattice scalar and vector $\varphi^4$ models \cite{cccppg,ccp},
in two and three dimensional XY models \cite{cccp}, in the $\Theta$-transition 
of homopolymeric chains \cite{cmp}, and - analytically - 
in the mean-field XY model \cite{Firpo}. Moreover, in the light of a Riemannian
geometrization of Hamiltonian dynamics --  where Lyapunov exponents are
related with 
average curvature properties of submanifolds of configuration space 
\cite{CFP,CLP,CCP} -- also the temperature dependence of abstract geometric
observables has been investigated. Among them the averages of curvature 
fluctuations (that enter the analytic formula for the largest Lyapunov exponent
\cite{CCP}) exhibit a cusp-like pattern. The peak coincides 
with the phase transition point. Qualitatively similar peaks of curvature 
fluctuations have been reproduced in abstract geometric models (families of 
surfaces of ${\Bbb R}^3$ where the variation of a parameter leads to a
change of the topological genus), whence the {\it heuristic} argument
about the topological meaning of the peaks of configuration-space-curvature
fluctuations at a phase transition. These fluctuations
 have been obtained as time
averages, computed along the dynamical trajectories of the Hamiltonian systems
under investigation. Now, time averages of geometric observables are usually
found in excellent agreement with ensemble averages 
\cite{cccp,cccppg,CCP,CasettiPettini} therefore one could argue that the 
mentioned singular-like patterns
of the averages of geometric observables are simply the precursors of truly
singular patterns due to the tendency of the measures of all the statistical 
ensembles to become singular in the limit $N\rightarrow\infty$ when a phase
transition is present. In other words, geometric observables, like any other
``honest'' observable, already at finite $N$ would feel the eventually
singular character of the statistical measures, and if this was the true
explanation we could not attribute 
the cusp-like patterns of curvature fluctuations to special geometric features 
of configuration space. Hence the motivation of the present paper. We aim 
at elucidating this important point by working out {\it purely geometric} 
informations about the submanifolds 
(to be specified below) of configuration space, {\it independently} of the 
statistical measures. 
In the present paper a new step forward is done in the direction
of supporting the {\it topological conjecture} mentioned above.

Our geometrical framework is the configuration space $M$ of systems whose
degrees of freedom are uncostrained and are real numbers, thus $M={\Bbb R}^N$.

Our statistical framework is the canonical ensemble whose
volume in $M$ is given by the configurational partition function 
$Z_C=\int \prod_{i=1}^N dq_i \exp [-V(q)]$, where $q=(q_1,\dots ,q_N)\in{\Bbb 
R}^N$. 
Finally, our topological framework is elementary Morse theory
\cite{Milnor,Hirsch}. Let us here recall its basic idea with the help of Fig.1.
The ``U--shaped'' cylinder of Fig.1 is the ambient manifold $M$ where a function
$V:M\mapsto{\Bbb R}$, smooth and bounded below, is defined to be the height of
any point of $M$ with respect to the floor-plane.
For any given value $u$ of
the function $V$ two kinds of submanifolds are determined: the level sets 
$\Sigma_u$ of all the points $x\in M$ for which $V(x)=u$, and $M_u$, the part 
of $M$ below the level $u$, i.e. the set of all points $x\in M$ such that
$V(x)\le u$.
The remarkable fact of Morse theory is that from the knowledge of all 
the critical points of $V$, i.e. those points where $\nabla V=0$, and of
their Morse indexes, i.e. the number of negative eigenvalues of 
the Hessian of $V$,
one can infer the topological structure of the manifolds $M_u$, 
provided that the critical points are nondegenerate, i.e. with
nonvanishing eigenvalues of the Hessian of $V$. 
Two such points are marked in Fig.1, at the bottom of $M$ and at some 
intermediate level $u_c$ for which $\Sigma_{u_c}$ is an eight-shaped curve.
It is evident from Fig.1 that the manifolds $M_{u<u_c}$ are {\it not
diffeomorphic} to the manifolds $M_{u>u_c}$: the formers are homeomorphic to
a disk and the latters are homeomorphic to a cylinder. The same happens (in
general) to the boundaries $\Sigma_u$ that here are circles for $u<u_c$ and 
become the topological union of two circles for $u>u_c$. This simple example 
displays the general fact that {\it passing a critical level of a Morse 
function is in one-to-one relation with a topology change}. A critical level is 
a surface $\Sigma_u$ that contains one or more critical points. 

Let us now consider the configuration space $M={\Bbb R}^N$ of a physical 
system and its potential $V$ as the Morse function. The interesting
things are supposed to occur below some large value ${\overline u}$
so that the corresponding (large) subset ${\overline M}\subset M$ is compact.
Then ${\overline M}$ and all its submanifolds $M_u$ are given a Riemannian 
metric $g$.
On all these manifolds $(M_u,g)$ there is a standard
invariant volume element: $d \eta =\sqrt{{\em det}(g)}dq^1\dots dq^N$. 

In order to study the topology of the family $\{(M_u,g)\}$ we should find,
analytically or numerically, all the critical points of $V$, but at large
$N$ this is a formidable task, therefore we have approached this problem
as follows. Generalizing a simple geometric
example reported in \cite{cccp,cccppg}, we have computed the total degree
of second variation $\sigma_K^2$
of the Gaussian curvature, i.e. $\sigma_K^2=\langle K_G^2\rangle_{\Sigma_u} -
\langle K_G\rangle_{\Sigma_u}^2$ where $\langle\cdot\rangle$ stands for
integration over the surface $\Sigma_u$, as a function of $u$ in the
neighborhood of a critical point. This is possible in general, independently
of the potential $V$, because any  
Morse function can be parametrized in the neighborhood of a critical point,
located at $x_0\in M$, by means of the so-called {\it Morse chart},
i.e. a system of local coordinates $\{x_i\}$ such that 
$V(x)=V(x_0)-\sum_{i=1}^kx_i^2+\sum_{i=k+1}^Nx_i^2$ ($k$ is the Morse index).
Then standard formulae for Gauss curvature of hypersurfaces of ${\Bbb R}^N$
\cite{thorpe} can be used to explicitly work out $K_G$ and $\sigma_K^2$.
 The intersection of an hypersphere of unit radius - centered around 
$u=0$ (the critical point) - with each $\Sigma_u$ is used to bound the domain 
of integration. 
The numerically tabulated results are reported in Fig.2 and show that
$\sigma_K^2$ develops a sharp, singular peak as the critical surface
is approached. It seems therefore reasonable to apply this geometric
probing of the presence of critical points, and hence of topology changes,
to the study of possible topology changes of the manifolds $(M_u, g)$.
In fact the manifolds $M_u$ inheritate - by somewhat smoothing them - the 
peculiar geometric properties of the $\{\Sigma_u\}_{u\leq u_c}$ -- due 
to the presence of critical points -- because of the 
relation $\int_{M_u}f d\eta =\int_0^u\,dv\int_{\Sigma_v}f\vert_{\Sigma_v} 
d\omega/\Vert\nabla V\Vert$, where $d\omega$ is the induced measure on 
$\Sigma_v$ and $f$ a generic  function \cite{federer}. 
The surface $\Sigma_{u_c}$ defined by $V(x)=u_c$ is a degenerate quadric,
therefore
close to it some of the principal curvatures \cite{thorpe} of the
surfaces $\Sigma_{u\simeq u_c}$ tend to
diverge. Such a divergence is generically detected by any function of the 
principal curvatures and thus, for
practical computational reasons, instead of Gauss curvature  (which is the 
product of all the principal curvatures)
we shall consider the total second variation of the {\it scalar} curvature 
${\cal R}$ (i.e. the sum of all the possible
products of two principal curvatures) of the manifolds $(M_u, g)$, according
to the definition 
\begin{eqnarray}
\sigma_{\cal R}^2(u) &= &[V\!ol(M_u)]^{-1} \int_{M_u}d\eta \left( {\cal R} - \varrho_u 
\right)^2 \\
\label{sigma2R} 
\varrho_u &= &[V\!ol(M_u)]^{-1} \int_{M_u}d\eta \, {\cal R} 
\label{aveR}
\end{eqnarray}
with ${\cal R}= g^{kj}R^l_{klj}$, where $R^l_{kij}$ is the 
Riemann curvature tensor \cite{doCarmo}, and $V\!ol(M_u)= \int_{M_u}d\eta$.

Let us now consider the family of submanifolds $(M_u, g)$
associated with the so-called $\varphi^4$ model, on a $d$-dimensional lattice 
${\Bbb Z}^d$, described by the potential function
\begin{equation}
V=\sum_{\alpha\in{\Bbb Z}^d}\left( - \frac{\mu^2}{2}
q_\alpha^2 + \frac{\lambda}{4!} q_\alpha^4\right) + 
\sum_{\langle\alpha\beta\rangle\in{\Bbb Z}^d}\frac{1}{2}J (q_\alpha -q_\beta )^2
\label{potfi4}
\end{equation} 
where $\langle\alpha\beta\rangle$ stands for nearest-neighboring sites. 
We consider
$d=1, 2$; this system has a discrete ${\Bbb Z}_2$-symmetry and short-range
interactions, therefore in $d=1$ there is
no phase transition whereas in $d=2$ there is a symmetry-breaking transition
(this system has the same universality class of the $2d$ Ising model).

The potential in Eq.(\ref{potfi4}) defines the subsets $M_u$ of configuration 
space, these subsets are given the structure of Riemannian manifolds $(M_u,g)$ 
by endowing all of them with the {\it same} metric tensor $g$. However,
as far as we want to learn something about the
topology of these manifolds, the choice of the metric $g$ is arbitrary.
We have therefore chosen three different types of metrics, one conformally
flat and the others non-conformal, according to a compromise between
simplicity and non-triviality. They are:
{\it i)} $g^{(1)}_{\mu\nu}=[A - V(q)]\delta_{\mu\nu}$, i.e. a conformal
deformation
of the euclidean flat metric $\delta_{\mu\nu}$, $A>0$ is an arbitrary constant 
larger than ${\overline{u}}$; {\it ii)}  $g^{(2)}_{\mu\nu}$ and 
$g^{(3)}_{\mu\nu}$ are non-conformal metrics defined by
\begin{equation}
(g_{\mu\nu}^{(k)})=\left(\matrix{f^{(k)}&0&1 \cr
                      0&{\Bbb I}&0 \cr
                      1&0&1 \cr}\right)~~,~~~~k=2,3
\label{metriche}
\end{equation}
where ${\Bbb I}$ is the $N-2$ dimensional identity matrix, 
$g^{(2)}$ is obtained by setting 
$f^{(2)}=\frac{1}{N}\sum_{\alpha\in{\Bbb Z}^d}q_\alpha^4 + A$, 
and $g^{(3)}$ by setting $f^{(3)}=\frac{1}{N}\sum_{\alpha\in{\Bbb Z}^d}
q_\alpha^6 + A$, with $A>0$, and $\alpha$ labels the $N$ lattice sites
of a linear chain ($d=1$) or of a square lattice ($d=2$, $N=n\times n$).
Simple algebra \cite{doCarmo} yields the scalar curvature 
function for each metric:
\begin{eqnarray}
{\cal R}^{(1)}&=&(N-1)\left[\frac{\triangle V}{(A-V)^2} - \frac{\Vert\nabla 
V\Vert^2}{(A-V)^3}\left(\frac{N}{4}-\frac{3}{2}\right)\right] 
\label{R1} \\
{\cal R}^{(k)}&=&\frac{1}{(f^{(k)}-1)}\left[ \frac{\Vert{\tilde\nabla} 
f^{(k)}\Vert^2}{2(f^{(k)}-1)} -{\tilde\triangle} f^{(k)} \right]~,~~~k=2,3 
\label{R23}
\end{eqnarray}
where $\nabla$ and $\triangle$ 
are euclidean gradient and laplacian respectively; $\tilde\nabla$
and $\tilde\triangle$ do not contain the derivatives $\partial /\partial 
q_\alpha$ with $\alpha =1$ ($d=1$) or $\alpha =(1,1)$ ($d=2$).

We constructed an ad-hoc MonteCarlo algorithm to sample the geometric 
measure $d\eta$ by means of the standard ``importance sampling'' method 
\cite{binder}, then we applied it to the computation of 
$\sigma_{\cal R}^2(u)$, given by Eq.(\ref{sigma2R}), for the one- and two-
dimensional lattice $\varphi^4$ model defined in (\ref{potfi4}) with
the following choice of the parameters: $\lambda=0.6,~\mu^2=2,~J=1$.
The values
of ${\cal R}$ are computed according to Eqs.(\ref{R1}) and (\ref{R23}). 
In order to
locate the phase transition that occurs in the two-dimensional case, we
have computed $\langle V\rangle$ {\it vs} $T$ by means of both MonteCarlo
averaging with the canonical configurational measure, and Hamiltonian dynamics
(by adding to $V$ a standard kinetic energy term).
In the latter case the temperature $T$ is given by the average kinetic
energy per degree of freedom, and $\langle V\rangle$ is obtained as
time average.
Fig.3 shows a perfect agreement between time and ensemble averages, thus
we worked out Fig.3 by computing $200$ time averages
because they converge much faster than ensemble ones. The phase transition
point is well visible at $u_c=\langle V\rangle\simeq 3.75$.

In Figs.4 and 5 we synoptically report the patterns of $\sigma_{\cal R}^2(u)$ for
the one and two dimensional cases obtained at different lattice size with
$g^{(1)}$ (Fig.4), and obtained at given lattice size with $g^{(2,3)}$
(Fig.5). Peaks of $\sigma_{\cal R}^2(u)$ appear at $u_c$ - the value of
$\langle V\rangle$ at the phase transition point - in the two-dimensional
case, whereas only monotonic patterns are found in the one-dimensional case,
where no phase transition is present. 

``Singular'', cuspy patterns of $\sigma_{\cal R}^2(u)$  
(with the meaning that such attributes can have for numerical results) 
are found {\it independently} of any possible statistical mechanical effect, 
and {\it independently} of 
the geometric structure given to the family $\{M_u\}$ by the metric
tensors chosen. Though within the well evident limits of numerical simulations
and of a limited choice of different metrics, our results suggest that
the ``singular'' patterns are most likely to have their origin at a deeper
level than the geometric one, i.e. at the topological level.
Hence the observed phenomenology strongly hints at the occurrence of some
{\it major} change in the topology of the configuration-space-submanifolds
$\{M_u\}$ in correspondence with a second-order phase transition. Finally,
the expression "{\it major} topology change" is to suggest that 
a change of the cohomological type of the $M_u$ - or $\Sigma_u$ - might well
be a necessary but not sufficient condition for a phase transition, and that
in any case some ``big'' change has to  occur.

\smallskip
It is a pleasure to thank E.G.D. Cohen, M. Rasetti and G. Vezzosi for their
continuous interest in our work and for useful comments and suggestions.

\begin{figure}
\caption{Illustration of the relationship between topology and critical points.
The U-shaped manifold $M$ is born at $P_0$. The level surfaces $\Sigma_u$ and 
the parts of $M$ below them - $M_u$ - change topology when $u$ exceeds the 
height of the critical point $P_1$.}
\label{fig1} 
\end{figure}

\begin{figure}
\caption{Variance of Gauss curvature {\it vs} $u$ close to a critical point.
$\sigma^{2/N}_K$ is reported because it is dimensionally homogeneous to the
scalar curvature. Here $N=dim(\Sigma_u)=100$, 
and Morse indexes are: $k=1,15,33,48$, represented by solid, dotted, dashed,  
long-dashed lines respectively. }
\label{fig2} 
\end{figure}

\begin{figure}
\caption{Average potential energy {\it vs} temperature for the $2d$ lattice
$\varphi^4$ model. Lattice size $N=20\times 20$. The solid line, made out of
200 points, refers to time averages. Full circles represent MonteCarlo
estimates of canonical ensemble averages. The dotted lines locate the phase
transition.}
\label{fig3} 
\end{figure}

\begin{figure}
\caption{Variance of the scalar curvature of $M_u$ {\it vs} $u$ computed with
the metric $g^{(1)}$. Full circles
correspond to the $1d$-$\varphi^4$ model with $N=400$. Open circles refer to
the $2d$-$\varphi^4$ model with $N=20\times 20$ lattice sites, and full 
triangles refer to $40\times 40$ lattice sites (whose values are rescaled for
graphic reasons). }
\label{fig4} 
\end{figure}

\begin{figure}
\caption{Variance of the scalar curvature of $M_u$ {\it vs} $u$ computed for
the $\varphi^4$ model with: metric $g^{(2)}$ in $1d$, $N=400$ (open triangles);
metric $g^{(2)}$ in $2d$, $N=20\times 20$ (full triangles);
metric $g^{(3)}$ in $1d$, $N=400$ (open circles); metric $g^{(3)}$ in $2d$, 
$N=20\times 20$ (full circles). }
\label{fig5} 
\end{figure}


\begin{references}
\bibitem{nota1} Here topology is meant in the sense of the de Rham's
cohomology.
\end{references}
\end{document}